\documentclass[11pt]{elsart}
\usepackage{natbib}
\usepackage{epsfig}
\usepackage{latexsym}
\usepackage{textcomp}
\usepackage[T1]{fontenc}
\usepackage{amsmath,amssymb}

\def\be{\begin{equation}}
\def\ee{\end{equation}}
\def\Z{{\bf Z}}
\newcommand{\ds}{{\sffamily DarkSUSY}}

\begin{document}

\begin{frontmatter}

\title{Photons from Dark Matter in a (non-Universal) Extra Dimension model}
\author{Marco Regis}
\ead{regis@sissa.it}
\address{{\small International School for Advanced Studies (SISSA/ISAS),
              Via Beirut 2-4, I-34014 Trieste, Italy and
              INFN, Sezione di Trieste, I-34014 Trieste, Italy}}

\begin{abstract}
We study the multi-wavelength signal induced by pairs annihilations at the Galactic center (GC) of a recently proposed dark matter (DM) candidate.
The weakly interacting massive particle (WIMP) candidate, named $A_-$, is the first Kaluza-Klein mode of a five dimensional Abelian gauge boson. Electroweak precision tests and the DM cosmological bound constrain its mass and pair annihilation rate in small ranges, leading to precise predictions of indirect signals from what concerns the particle physics side. The related multi-wavelength emission is expected to be faint, unless a significant enhancement of the DM density is present at the GC. We find that in this case, and depending on few additional assumptions, the next generation of gamma-ray and wide-field radio observations can test the model, possibly even with the detection of the induced monochromatic gamma-ray emission.
\end{abstract}

\begin{keyword}
Extra Dimension, Dark Matter, Indirect Detection, Galactic Center
\PACS{11.10.Kk, 12.60.-i, 95.35.+d, 98.35.Jk}
\end{keyword}

\end{frontmatter}

In the last decades gravitational evidences for dark matter (DM), based on different observables, such as rotation curves, velocity dispersions, gravitational lensing, large scale structure maps and cosmic microwave background (CMB) anisotropies have been accumulated at the galactic, cluster and cosmological scales.
Still, the identification of the DM component remains one of the most challenging issue of the physics today  (for recent reviews see~\cite{review1,review2}).
Weakly interacting massive particles (WIMPs) are a well motivated class of candidates for the non--baryonic component. The WIMP paradigm is well--known: In thermal equilibrium in the primordial bath, WIMPs decouple in the non--relativistic regime and the weak interaction leads the relic abundance to be of the order of the mean energy density of DM in the Universe today. 
Being (weak) interacting particles, WIMPs can annihilate in pairs in astrophysical structures, inducing detectable signatures, such as antimatter and neutrino fluxes, and multi--wavelength spectra.
Complementary to direct DM searches and to collider experiments testing extensions to the standard model (SM) of particle physics embedding a WIMP candidate, indirect detection can provide crucial information about the fundamental nature of DM.

\begin{figure}[t]
 \begin{minipage}[htb]{6.2cm}
   \centering
   \includegraphics[width=6.2cm]{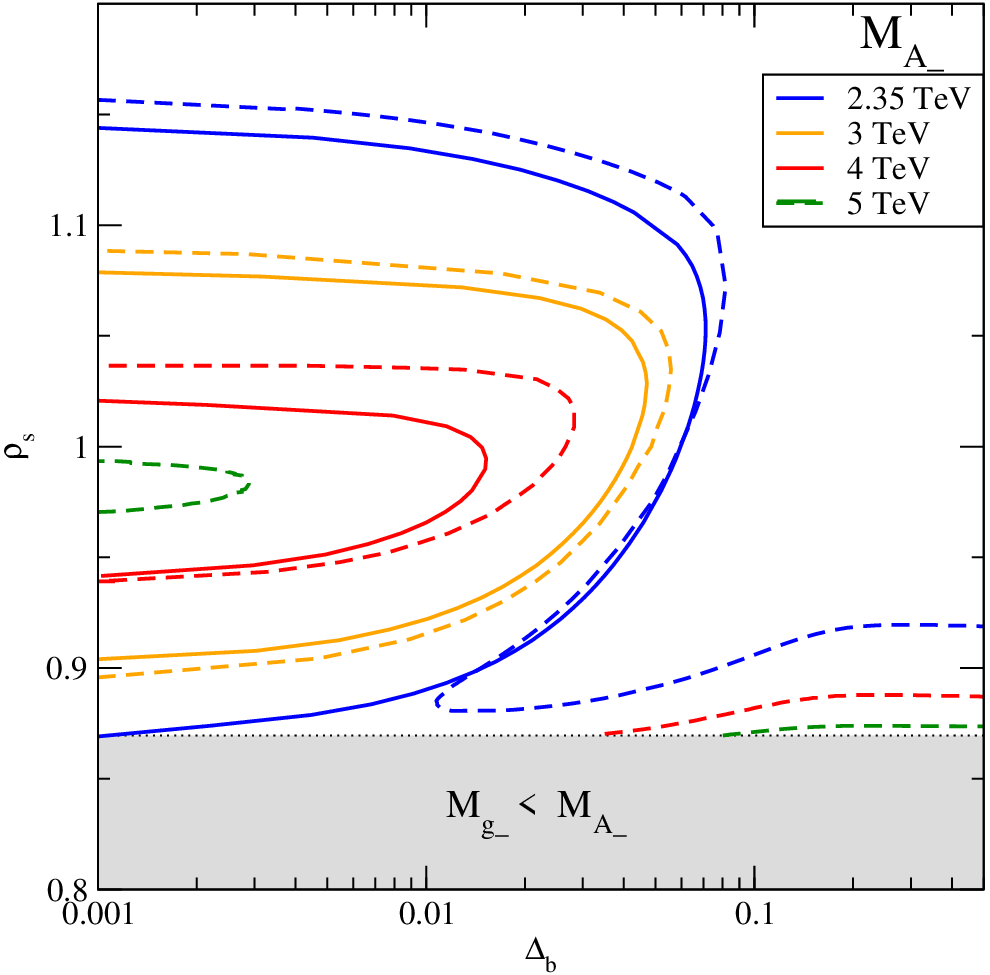}
 \end{minipage}
 \ \hspace{2mm}  \
 \begin{minipage}[htb]{6.2cm}
  \centering
   \includegraphics[width=6.2cm]{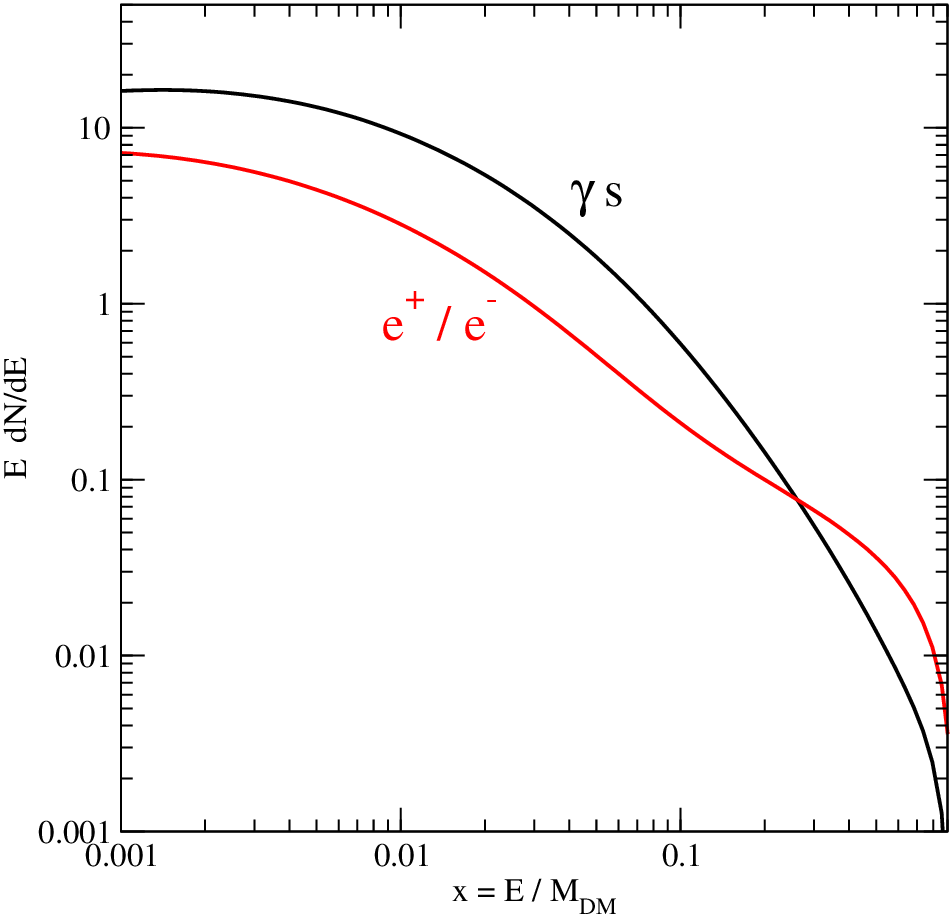}
 \end{minipage}
   \caption{{\small {\it Left Panel}: For a few selected values of the DM mass, isolevel curves of the parameter space ($\Delta_b = (M_{b_-}-M_{A_-})/M_{A_-}$, $\rho_s$) in which the $A_-$ relic abundance matches the best fit value $\Omega_{DM} h^2$ from
   cosmological observations. The solid and dashed lines refer to the two different setups described in the text. {\it Right Panel}: Gamma-ray and electron/positron differential spectra per annihilation of the DM candidate $A_-$ in the minimal framework.}}
   \label{fig1}
\end{figure}

In Ref.~\cite{Regis:2006hc} a new viable DM candidate was introduced. It is embedded in a flat five dimensional (5D) model compactified on an orbifold $S^1/\Z_2$, with gauge-Higgs unification and explicit Lorentz symmetry breaking in the bulk~\cite{Scrucca:2003ra,Panico:2005dh,Panico:2006em}.
The DM candidate, named $A_-$, is the first Kaluza--Klein (KK) mode of an Abelian 5D antiperiodic gauge boson field. It is the lightest KK particle odd under the ``mirror symmetry'', a discrete $\Z_2$ symmetry introduced in Ref.~\cite{Panico:2006em} to improve the naturalness of the model, namely to reduce the fine-tuning needed to stabilize the electroweak (EW) scale. 
In a relevant fraction of the parameter space, the DM candidate is nearly degenerate 
with the lightest antiperiodic fermion $b_-$ and the antiperiodic gluon $g_-$ (see the mass spectrum in Fig.~1 of~\cite{Regis:2006hc}). The mass for the first KK mode of antiperiodic fields is given at tree level by: 
\be
M_{g_-} =\frac{\rho_s }{2R}\;,\;\;\;\;
M_{A_-} = \frac{\rho}{2R}\;,\;\;\;\;
M_{f_-} = \sqrt{m_f^2 + \left(\frac{k_f}{2R}\right)^2}\;,
\label{eq:mass}
\ee 
where $f_-$ denotes an antiperiodic fermion, $R$ is the compactification radius of the covering circle $S^1$, $\rho_i$ and $k_f$ are the Lorentz breaking parameters of gauge boson and fermion respectively, and $m_f$ is the fermionic bulk mass.
EW bounds force the compactification scale of the model in the Multi--TeV regime~\cite{Panico:2006em}, leading to a lower bound for the DM mass: $M_{A_-}\gtrsim2.35$ TeV, assuming $\rho \sim 1$.
The $A_-$ pair annihilation rate is quite small ($\sigma_{ann} v \lesssim 5 \cdot 10^{-28} cm^3 s^{-1}$) compared to WIMPs in more standard scenarios (e.g. the lightest neutralino in supersymmetry and the $B^1$ in universal extra dimension (UED) scenario, see Ref.~\cite{review1});
however, being $b_-$ and $g_-$ strongly interacting particles, coannihilation effects greatly
enhance the effective annihilation rate, leading to relic abundances allowed by cosmological observations.
The curves in Fig.~\ref{fig1}a show the mass splitting between the DM candidate and the coannihilating particles, such that the $A_-$ relic density matches the cosmological amount of DM today~\cite{Spergel:2006hy}. They are expressed in terms of ($\Delta_b, \rho_s$) with $\Delta_b \equiv (M_{b_-}-M_{A_-})/M_{A_-}$, taking into account radiative corrections as described in Ref.~\cite{Regis:2006hc}.
Fig.~\ref{fig1}a shows two different setups. Indeed, the gauge group of the model is $G \times G'$,
where $G=SU(3)_w\times SU(3)_s\times U(1)_1$ and $G' = SU(3)'_{s}\times U(1)'$ as in the first construction of Ref.~\cite{Panico:2006em} (dashed lines) or $G' = U(1)'$ as in the framework (in the following called ``minimal'' framework) considered in Ref.~\cite{Regis:2006hc} (solid lines); antiperiodic gluons are present only in the first setup and differences between the two cases for what concerns the dark matter relic density computation are described in Ref.~\cite{Regis:2006hc}.
Combining EW and cosmological bounds, the $A_-$ mass is constrained in the narrow window 2.35~-~5 TeV. 
Naturalness arguments can restrict even more the parameter space of the model. 
Indeed the value for the mass preferred by EW observables is $\sim 3$ TeV~\cite{Panico:2006em} and the fine tuning associated to the DM relic density is minimized by the minimal framework~\cite{Regis:2006hc}, where $A_-$ annihilates only with $b_-$, leading to $(M_{b_-}-M_{A_-})/M_{A_-}\lesssim 7\%$ (see Fig.~\ref{fig1}a).

Couplings with SM fermions are highly suppressed\footnote{This fact implies a very small elastic scattering cross section between $A_-$ and light quarks, and the expected direct DM signals are well below the sensitivity of current detection experiments.}
 since the latter (with the exception of bottom and top quarks) are mainly
localized on the 4D brane at $y=0$, where the $A_-$ wavefunction vanishes, being antiperiodic on $S^1$.
As we can see from the mass spectrum in Ref.~\cite{Regis:2006hc}, some
non--standard states are energetically accessible by the $A_-$ pair annihilation.
They belong to SU(2)$_w$ singlet, doublet or triplet representation and they are the first massive eigenstates obtained from the diagonalization of the Lagrangian describing the coupling between 5D periodic fermions and 4D localized fermions. Non--SM singlets and doublets are the KK excitations of the SM fermions and they are mainly localized.
The case of triplets is different: they are not coupled with boundary fields and
their wavefunctions are constant in the bulk, hence largely overlapping the $A_-$ wavefunction.
Moreover the associated number of degrees of freedom is huge and so
the dominant final states of the $A_-$ annihilation cross section are triplets
fermions.
More precisely the annihilation branching ratios are: $75\%$ into $b_+ \bar b_+$, $24\%$ into $\tau_+\bar \tau_+$ (with $b_+$ and $\tau_+$ being the $SU(3)_w$ triplet of the bottom and tau tower, respectively) and
$1\%$ into SM quarks. The subsequent decays of $b_+$ and $\tau_+$ generate quark pairs ($38\%$), $\tau$ lepton pairs ($6\%$) and neutrinos ($6\%$), charged ($25\%$) and neutral ($12\%$) weak gauge bosons, and Higgs bosons ($12\%$).

Photons cannot be directly produced by WIMP pair annihilations at tree level. A continuum photon spectrum is generated in cascades with hadronization of unstable two-body annihilation final states into $\pi^0$s and their subsequent decays.
The energy of these photons is in the $\gamma$-ray band.
Electrons and positrons can be directly or indirectly produced in WIMP pair annihilations as well. The associated radiative processes can act as source for a multi--wavelength spectrum covering from radio to soft--gamma ray frequencies.
In Fig.~\ref{fig1}b we show the differential energy spectra per $A_-$--annihilation into $\gamma$-rays and $e^+-e^-$ in the minimal DM framework (variations of the DM mass within the allowed range do not affect the spectra in a sizable way).
In the first case, on top of the spectrum originated from $\pi^0$ decay, we consider the contribution of primary gamma-rays from final state radiation following the line of Refs.~\cite{Bergstrom:2004cy} and \cite{Birkedal:2005ep}.
We derive the differential yields through simulations of decay and hadronization performed with the PYTHIA Monte--Carlo code~\cite{Sjostrand:1993yb}.
The two spectra are soft since quarks and gauge bosons are the dominant annihilation modes. From the point of view of indirect searches, this feature distinguishes $A_-$ from the UED WIMP candidate $B^1$~\cite{Servant:2002aq,Cheng:2002ej}, whose pairs annihilation branching ratios are dominated by charged leptons and harder spectra are produced. 
The electron/positron and gamma-ray yields of Fig.~\ref{fig1}b are at a comparable level, being mostly generated by the production and decays of charged and neutral pions, respectively, with the two chains having analogous efficiency.

\begin{figure}[t]
 \begin{minipage}[htb]{6.2cm}
   \centering
   \includegraphics[width=6.2cm]{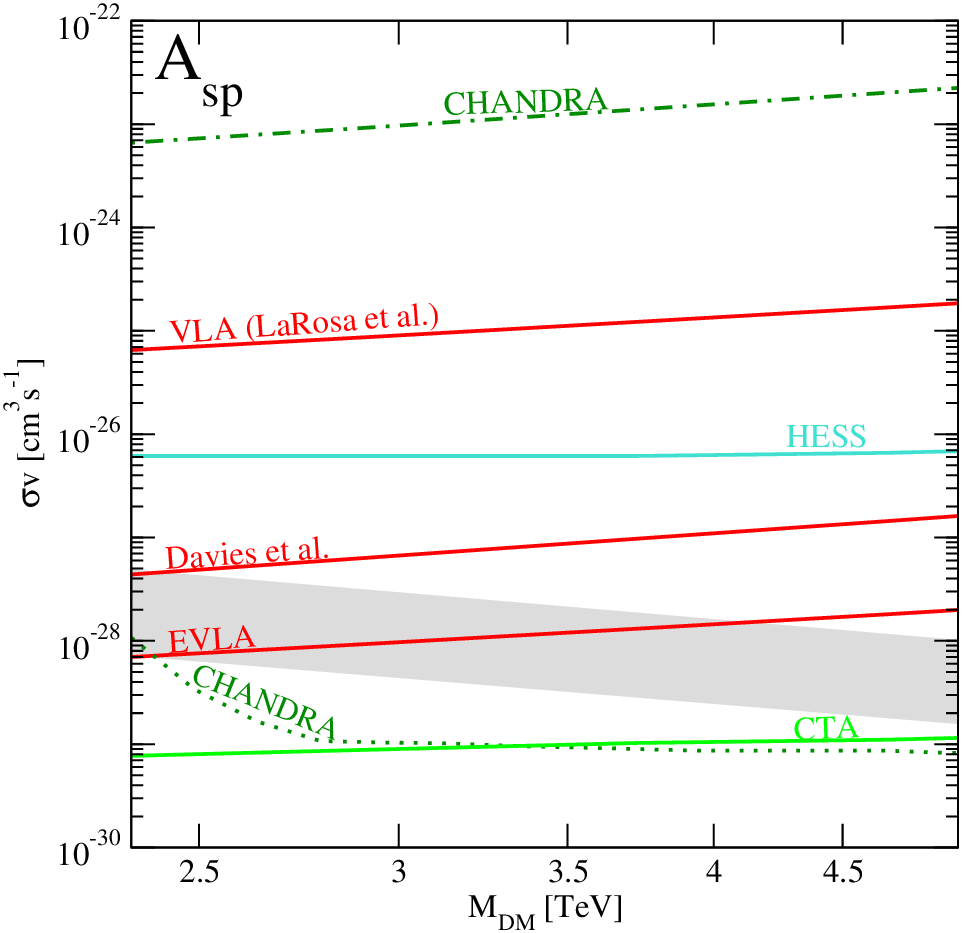}
 \end{minipage}
 \ \hspace{2mm}  \
 \begin{minipage}[htb]{6.2cm}
  \centering
   \includegraphics[width=6.2cm]{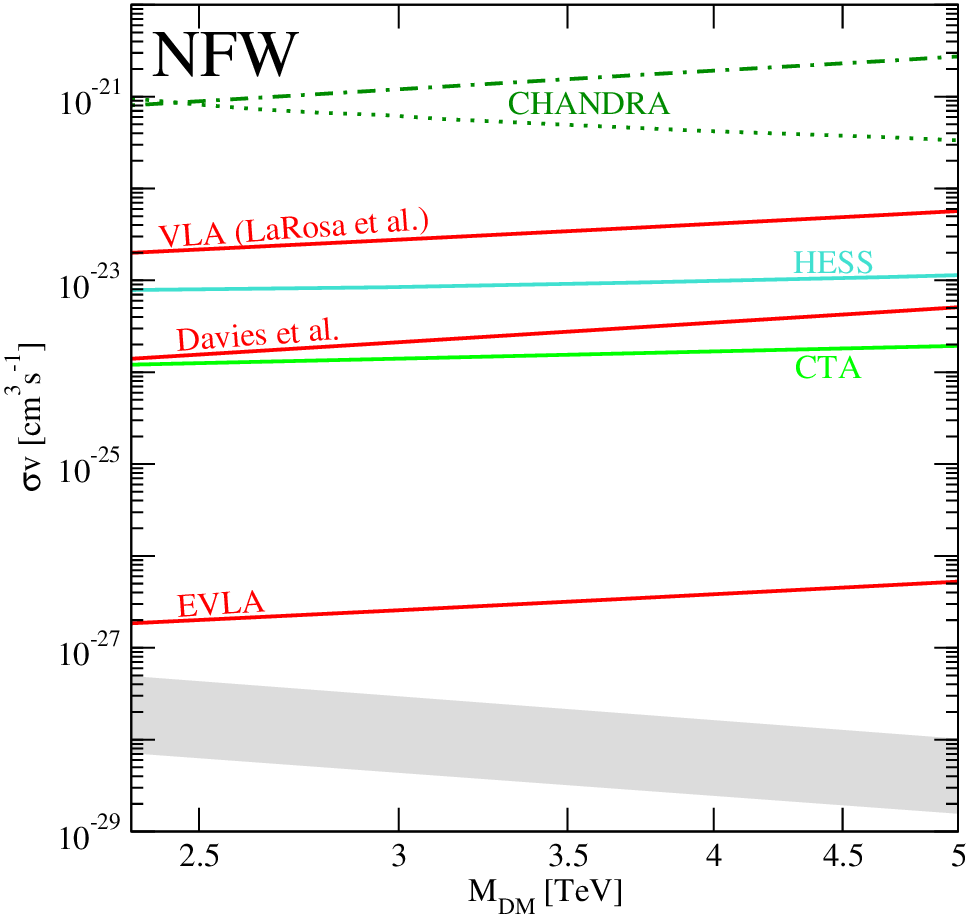}
 \end{minipage}
   \label{fig:svvsm}
   \caption{{\small Exclusion limits on the $A_-$ annihilation cross section as a function of the WIMP mass. The {\it Left} and {\it Right Panels} show the cases of $A_{sp}$ and NFW profiles, respectively.}}
\end{figure}

As listed in Table~2 of Ref.~\cite{Regis:2006hc}, all the $A_-$-annihilation processes occur through t or u-channels mediated by an antiperiodic fermion. At a given DM mass, the only free parameter affecting in a sizable way the cross section computation is the mass of the mediator.
As already mentioned before, in the minimal DM framework, the relic abundance is driven mostly by the $b_-$ coannihilation, highly constraining $M_{b_-}$ and hence $k_b$ (see Eq.~\ref{eq:mass}). It leads the total $A_-$ annihilation cross section within a small range, since the triplet pairs associated to the $b$--multiplet are the dominant annihilation modes.
The 5D Lorentz symmetry breaking was introduced to achieve the correct value for the top mass; the Lorentz breaking parameter $k_i$ associated to other fermions can be safely taken $\sim1$. For our purposes $k_b$ and $k_{\tau}$ are relevant in the computation of the annihilation cross section in the non--minimal scenario, where we assume $k_i\lesssim2$. The allowed total annihilation cross sections as a function of the WIMP mass are shown in Fig.~\ref{fig:svvsm} by the filled band; in the minimal framework this region shrinks to its upper boundary.

The WIMP candidate we consider has a small annihilation rate and a quite heavy mass.
Looking for a WIMP induced signal might seem hopeless, unless
we concentrate on a region where the DM density is very large and where emissions from other astrophysical sources are faint.
In the following we focus on photon emissions at the Galactic Center (GC), where the first condition is definitely fulfilled, while the second is not completely satisfied, at least in the wide range of frequency over which a source labeled Sgr~A$^*$ has been detected. On other hand, the luminosity associated to this source turns out to be quite low, at a level comparable to the expected DM signal.
 
In the determination of the $A_-$-induced emission, the ingredients related to the particle physics side are quite strictly constrained, while the astrophysical uncertainties remain large, as for any WIMP model.
The Milky Way DM halo profile is not completely established, in particular for what concerns its inner region. As well known, there is some tension between N--body simulation results and observations of spiral galaxies about the central cusp/core.
Sufficiently bright DM signals, at the level needed to test the WIMP model under investigation, appear only for halo profiles with a large overdensity in the GC region.
We follow the analysis in Ref.~\cite{Bertone:2005hw} and focus our attention on a distributions obtained from the evolution of a Navarro-Frenk-White (NFW) profile~\cite{Navarro:1996gj}, including the deepening in the Galactic potential well generated by the slow adiabatic formation of the supermassive black hole (SMBH) and of the stellar component in the inner Galaxy. 
In such profile, hereafter labeled $A_{sp}$, the effect of self annihilation triggers the density in the innermost region. The numerical profiles were kindly provided by the authors of Ref.~~\cite{Bertone:2005hw} for values of $(\sigma v)$ and $M_{DM}$ in the filled band in Fig.~\ref{fig:svvsm}, while we derive scaling laws in the rest of the parameter space starting from results reported again in Ref.~\cite{Bertone:2005hw}.

In the GC case, synchrotron radiation and, to a smaller extent, inverse Compton scattering on CMB and starlight are the most significant radiation mechanisms. The estimation of the associated emissions requires a description of the electrons/positrons propagation and of their energy transfers after production in WIMP annihilations. We need a model of the galactic medium, to fix the diffusion coefficient, the magnetic field, the advective/convective velocity and the absorption effects. For this treatment we refer to Ref.~\cite{Regis:2008ij}.

The compact radio source Sgr~A$^*$ associated to the SMBH has been detected at the GC (see the catalog in Ref.~\cite{Narayan:1997ku} and reference therein), together with its infrared~\cite{Genzel:2003as} and X-ray~\cite{Baganoff:2001ju} counterparts. A gamma-ray emission from the GC region has been detected as well~\cite{MayerHasselwander:1998hg,Aharonian:2004wa}, but with experimental angular resolutions not sufficient to identify the source and its precise location.
Any of the portions of such multi--wavelength spectrum turns out to be incompatible with emissions induced by WIMP annihilations and the detected signals will be exploited here to derive upper limits.
Also diffuse emissions from the inner region of the Galaxy has been detected
at different frequency bands,
and, in case of shallow DM halo profile, can severely constrain WIMP models.
The procedures implemented to extract the limits shown in Figs.~\ref{fig:svvsm}a and \ref{fig:svvsm}b were outlined in Ref.~\cite{Regis:2008ij}. 
We assume $A_-$ accounting for the whole DM content of the Universe and all the numerical calculations are performed with the help of the \ds~package~\cite{Gondolo:2004sc}.
Together with bounds associated to the mostly investigated profile in this Letter, i.e. the $A_{sp}$ profile, we compute, for comparison, limits on the WIMP parameter space in case of a NFW profile, namely the mostly investigated case in the literature.
We plot the tightest bounds in gamma-ray and radio bands obtained from spectral and angular analysis, comparing the WIMP signals with the emission detected by the $\gamma$-ray air Cherenkov Telescope (ACT) HESS~\cite{Aharonian:2004wa,Aharonian:2006au} and with upper bounds in the radio surveys of Refs.~\cite{davies76} and \cite{LaRosa:2000}.
In the X-ray band, synchrotron emission would require very strong magnetic field, especially in case of soft electron/positron spectrum. This could be possible only in the innermost region of the Galaxy, depending on the model considered for accretion flow around SMBH, hence the size of the DM induced source is very small. Limits on WIMP parameter space can be extracted by the comparison with the Sgr~A$^*$ emission detected by the Chandra observatory~\cite{Baganoff:2001ju}, but they are highly model dependent. We plot the weakest constraint among the three cases with different choice of magnetic field radial profile of Ref.~\cite{Regis:2008ij}.
The angular size of the emission induced by the inverse Compton scattering on CMB is much larger and the signature estimate involves more reliable assumptions on the magnetic field strength at larger scales.
The limit extracted by the comparison with the detected X-ray diffuse emission~\cite{Muno:2004bs} (dashed-dotted lines) is much less constraining (but more robust) with respect to the limit associated to the point--like synchrotron source (dotted lines); the fact that the latter is excluding the whole $A_-$ parameter space in the $A_{sp}$ case should not be overemphasized, given the critical extrapolations involved in this result.

Then we derive projected limits from forthcoming gamma-ray surveys and wide-field radio observations.
For heavy WIMP models, the parameter space can be more efficiently studied by ACTs rather than space satellites, due to the different energy ranges of detection. We consider the next generation of ACT, under development by the Cherenkov Telescope Array (CTA) project, and scheduled for 2013, assuming performances as outlined in~\cite{CTA:2007}.
A diffuse radio emission was reported both in the Milky Way atlas of Ref.~\cite{Haslam:1982} and in the GC image of Ref.~\cite{LaRosa:2005ai}. However, the two surveys have too poor angular resolutions to resolve the spatial profile of the emission in the innermost region. In the GC map of Ref.~\cite{LaRosa:2000}, such emission does not seem completely isotropic and tight constraints are derived from patches of the map with no astrophysical background.
The radio projected limits plotted in Figs.~\ref{fig:svvsm}a and \ref{fig:svvsm}b are extracted again following Ref.~\cite{Regis:2008ij}, but assuming a detector sensitivity improved by a factor 10, as proposed in the EVLA project~\cite{EVLA}.

The scale at which the formation of the SMBH could have influenced the DM distribution is far below the resolution of both numerical simulations and observations.
The related DM spike in the $A_{sp}$ profile greatly enhances signals in the innermost region of the GC and the comparison with the Sgr~A$^*$ source is very constraining, especially for $(\sigma v)/M_{DM}\gtrsim 10^{-32} cm^{-3}s^{-1} GeV^{-1}$~\cite{Bertone:2005hw}.
The limits associated to diffuse emissions are less constraining, since involve angular scales where the enhancement in the DM distribution $A_{sp}$ is less pronounced with respect to an NFW profile, being related to the deepening in the potential well induced only by the stellar component.
For the same reason, being the DM induced radio source more extended than the DM source itself, and thus than the gamma-ray source, the bound associated to wide field radio signal is less stringent with respect to gamma-ray limit in case of $A_{sp}$ profile. The picture is reversed for the NFW distribution.
In case of $A_{sp}$ profile, all the multi-wavelength constraints extracted from past surveys, excluding the synchrotron X-ray bound, do not limit the region allowed by cosmological and EW bounds (filled band).
On the other hand, in the next decade, the model could be completely tested through its gamma-ray emission by the CTA experiment.
The plotted exclusion curve is computed assuming an effective area A$_{eff}=1$ km$^2$ and an exposure time $t_{exp}= 250$ hours in $5$ years of collecting data.
Depending on the properties of the galactic radio diffuse emission at small scales, the EVLA project could test the $A_-$ radio profile in a large fraction of the parameter space, covering basically the whole region of the minimal DM framework.
In the case of NFW profile, no significant constraint can be derived. Note however that radio wide field observations can be much more efficient than gamma-ray measurements.

\begin{figure}[t]
 \begin{minipage}[htb]{6.2cm}
   \centering
   \includegraphics[width=6.2cm]{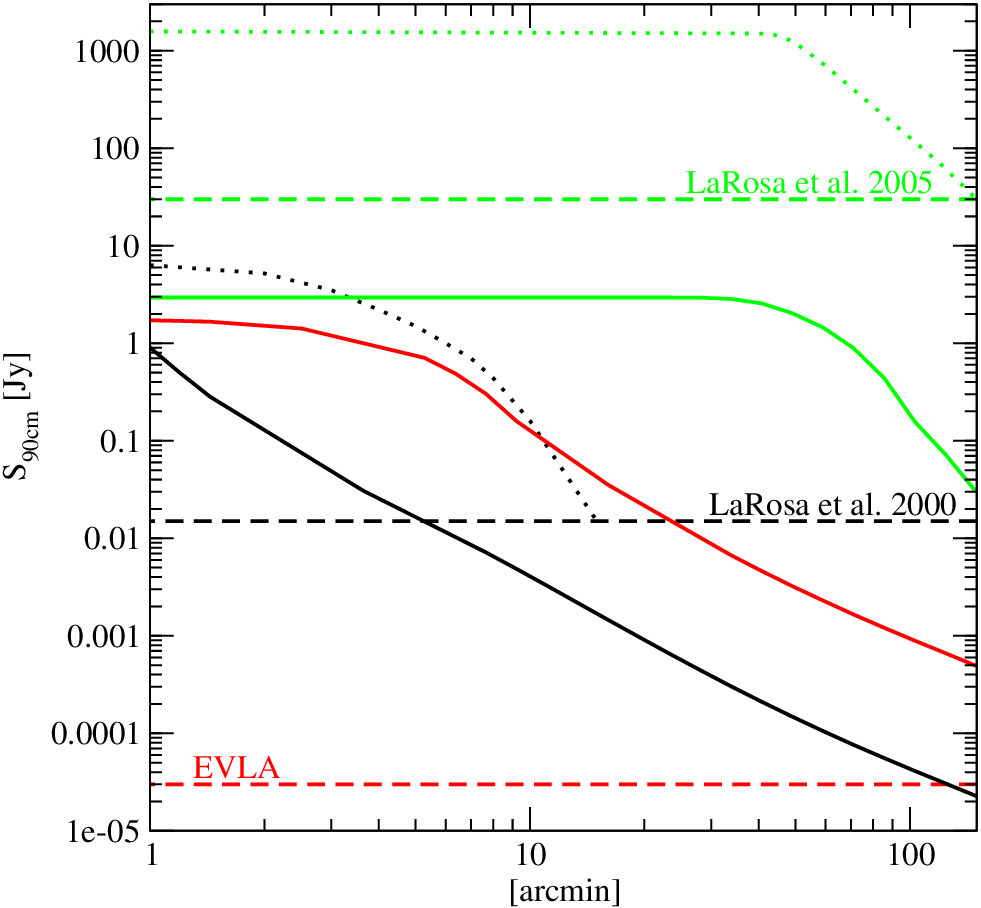}
 \end{minipage}
 \ \hspace{2mm}  \
 \begin{minipage}[htb]{6.2cm}
  \centering
   \includegraphics[width=6.2cm]{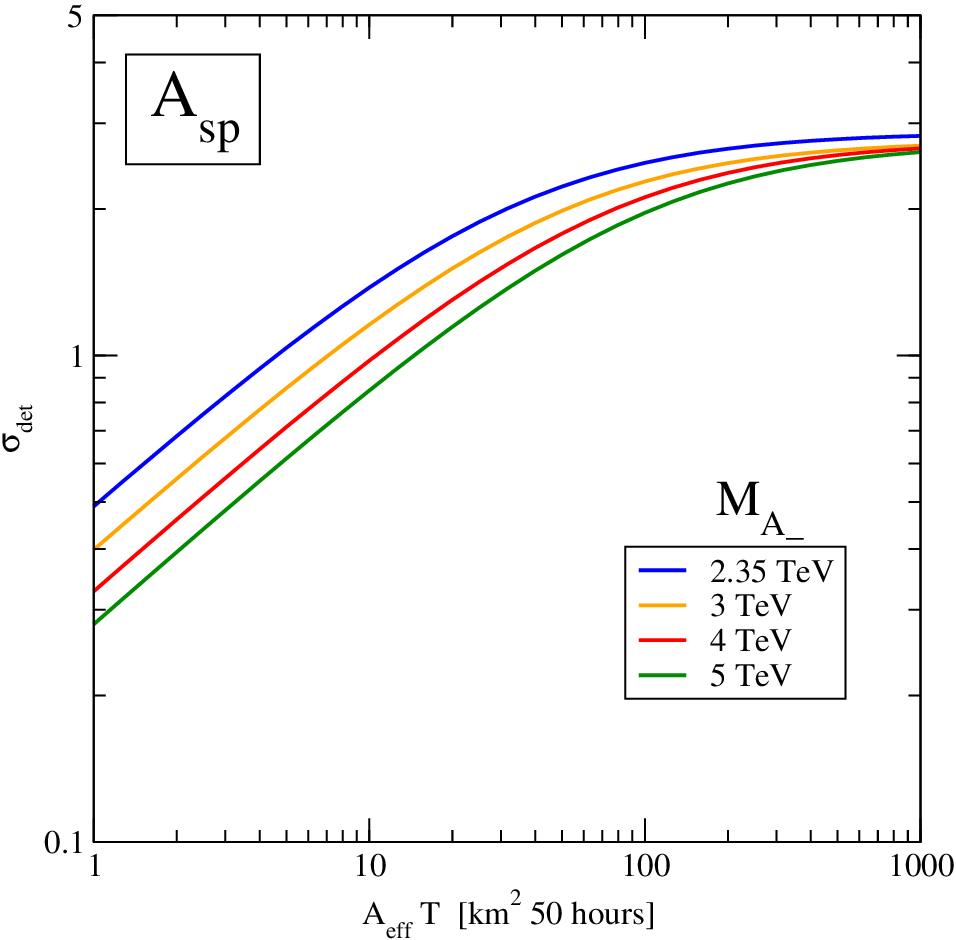}
 \end{minipage}
   \caption{{\small {\it Left Panel}: Angular profiles of the expected DM induced synchrotron source (solid lines) and of the detected diffuse emissions (dotted lines) at 90 cm in the surveys of Refs.~\cite{LaRosa:2000} (green) and \cite{LaRosa:2005ai} (black). The DM signal profile is shown also for a hypothetical EVLA observation with FWHM=200\textquotedbl (red). We consider as benchmark case the $A_{sp}$ halo profile, $M_{DM}=3$ TeV and $\sigma_{ann} v = 3 \cdot 10^{-28} cm^3 s^{-1}$. Dashed lines show the experimental sensitivities. {\it Right Panel}: For a few selected values of the DM mass, detectability of a monochromatic gamma-ray signature by the CTA project as a function of effective area $\times$ exposure time. The latter is expressed in terms of 1 km$^2 \times$ 50 hours, which can be considered as a conservative estimate for one year of observation by CTA.}}
   \label{fig3}
\end{figure}

Radio observations with a wide field of view have detected extended emissions from the GC region. In Fig.~\ref{fig3}a we plot schematic representations of the angular shape of the signals at 90 cm, as detected in the map of Ref.~\cite{LaRosa:2000} (FWHM=43\textquotedbl) and  Ref.~\cite{LaRosa:2005ai} (FWHM=40\textquotesingle). For both we sketch the profile of the extended source along its longitudinal axis.
The level of the DM induced emission filtered over the same experimental angular resolutions is also shown, together with the $3\sigma$ sensitivity of the detectors. We take as benchmark case for the $A_-$ candidate, a mass of 3 TeV and an annihilation rate of $\sigma_{ann} v = 3 \cdot 10^{-28} cm^3 s^{-1}$. The DM distribution considered is again the $A_{sp}$ profile.
If the astrophysical radio diffuse emission is approximately isotropic at any scales, bounds on WIMP parameter space that could be extracted are not so stringent, as shown by the green curves, which is averaged over an angular resolution of 40 arcmin. On the other hand, if, on smaller scales, regions without contamination from astrophysical background are present, this type of surveys seems to be very promising, as shown in particular by the red curves, representing a hypothetical observation by EVLA with FWHM=200\textquotedbl. However, this picture is probably based on a too optimistic assumption and it has to be considered as a limiting case.

So far we have considered only continuum energy spectra of photons and electrons/positrons. 
The coupling between $A_-$ and electrons is very tiny, since the latter are completely localized on the 4D brane at the boundary of the extra dimension. Thus, for our purposes, the prompt emission in monochromatic electrons and positrons can be neglected.
Even if a gamma-ray continuum signal from DM annihilation exceeds the astrophysical
background, the identification of the DM component, which involves the exclusion of
any other astrophysical explanation, could be a difficult task.
The real ``smoking gun'' of a DM induced gamma-ray signal would be a monoenergetic
spectral signature.
By definition, the DM coupling with photons is highly constrained, but a direct
WIMP annihilation into $\gamma \gamma$ at one-loop level is allowed, producing photons with energy
$E_{\gamma}\simeq M_{DM}$, being WIMPs non-relativistic.
Since $A_-$ is an Abelian gauge boson, this process can occur through fermionic
boxes. The main contribution is given by fermion triplets in the loop, for the same reason 
(i.e. the delocalization) stated above referring to the tree level annihilation into fermions.
The cross section computation is performed following 
Ref.~\cite{Bergstrom:2004nr}, and obtaining $\sigma_{\gamma \gamma}v\simeq 2\cdot 10^{-4}\,\sigma_{ann}v$.
The total number of events associated to DM annihilations into monochromatic $\gamma \gamma$ in a detector pointing to the GC direction with angular resolution $\Delta\Omega$, is given by:
\be
N_{line}= 1.9\,\,10^{-13} \frac{\sigma_{line}v}{10^{-31} cm^3 s^{-1}}\Big(\frac{{\rm TeV}}{M_{DM}}\Big)^{2} \bar J(\Delta\Omega)\Delta\Omega \frac{A_{eff}}{m^2} \frac{T}{s}\label{eqgaline}\;.
\ee
The quantity $\bar J(\Delta\Omega)$, containing all the spatial information, is defined as:
\be
\bar J(\Delta\Omega) =\frac{1}{8.5\,{\rm kpc}}\Big(\frac{1}{0.3\,{\rm GeV/cm^3}}\Big)^2 \frac{2\pi}{\Delta\Omega}\int d\theta \exp{\Big(-\frac{\tan^2 \theta}{2\tan^2 \theta_d}\Big)}\int_{l.o.s.}\rho^{2}(l)dl \label{Jprof}
\ee
where $\rho$ is the
DM halo profile, $\theta$ is the angular off-set with respect to the GC and $l$ is the coordinate along the line of sight. In Eq.~\ref{Jprof} we consider a circular Gaussian angular resolution of width $\theta_d$ for the detector.
The ratio between the gamma-ray signals originated in an $A_{sp}$ and an NFW profiles is given by the ratio: 
$b = \bar J_{A_{sp}}(10^{-5}{\rm sr})/\bar J_{NFW}(10^{-5}{\rm sr})$, assuming $\Delta\Omega=10^{-5}$ sr for modern ACTs. In the range of mass and cross section of the $A_-$ model, it approximately follows the law:
$b\simeq 10^{4}\left[\left(\sigma_{ann}v/10^{-28}{\rm cm^3 s^{-1}}\right)\left({\rm TeV}/M_{DM}\right)\right]^{-0.8}$.
The dependence from the ratio $\sigma_{ann}v/M_{DM}$ reflects the fact that the initial DM distribution, from which the $A_{sp}$ profile is derived, has a spike around the SMBH. In this case, self-annihilations frequently occur in the innermost region, triggering the final shape.

The number of events associated to the $\gamma$-ray continuum background in a CTA bin can be obtained integrating the spectrum of the detected GC source and of the misidentified showers from hadrons and electrons~\cite{Bergstrom:1997fj} over an energy resolution of $10\%$.
The probability of disentangling $N_{l}$ events associated to the DM induced gamma-ray line from $N_{bg}$ events of the continuum background is related to $\sigma_{det}=N_{l}/\sqrt{N_{bg}+\epsilon_{sys}^2 N_{bg}^2}$, where $\epsilon_{sys}$ gives the level of systematic errors, taken to be $1\%$ for CTA~\cite{CTA:2007}.
We estimate $N_l$ to be a fraction $\epsilon_{DM}\sim 2.7\%$ of the total number of events. At fixed systematic error, the maximal significance which can be achieved increasing the effective area or the exposure time is $\sigma_{det}^{max}=\epsilon_{DM}/\epsilon_{sys}$, i.e. the plateau in Fig.~\ref{fig3}b.
A conservative guess for A$_{eff} \times$T$_{exp}$ is 1 km$^2 \times$ 50 hours in one year of observation by CTA. As shown in Fig.~\ref{fig3}b, the prompt monochromatic emission of $\gamma \gamma$ originated from $A_-$ annihilation in an $A_{sp}$ halo profile needs an extra factor of 100 in A$_{eff}\times$T$_{exp}$ in order to be detected at $\sim3\sigma$; this could be reached only with a quite larger setup than the minimal designed and in several years of observation.

To conclude, in this Letter we have outlined the properties of a DM candidate recently proposed, sketching its multi--wavelength indirect signal from the GC.
Cosmology and EW precision tests fix its mass and total annihilation cross section in a narrow window, which is compatible with the bounds associated to the detected emissions at the GC, but can be definitely tested by the forthcoming gamma-ray and wide-field radio surveys, if the Milky Way halo profile is spiky. We also discuss the possible detection of an induced gamma-ray line in the same framework.
On the other hand, in case of NFW or more shallow profiles, the model cannot be constrained through this multi--wavelength strategy.

We would like to thank G.~Bertone and D.~ Merritt for kindly providing some of DM halo profiles which have been used in this analysis. We also would like to thank M.~Serone and P.~Ullio for useful discussions.

\end{document}